# An Assessment of Weight-Length Relationships for Muskellunge, Northern Pike, and Chain Pickerel In Carlander's Handbook of Freshwater Fishery Biology

Joshua Daviscourt, Joshua Huertas,  and Michael Courtney
U.S. Air Force Academy, 2354 Fairchild Drive, USAF Academy, CO 80840
Michael.Courtney@usafa.edu

**Abstract:**  Carlander's Handbook of Freshwater Fishery Biology (1969) contains life history data from many species of freshwater fish found in North America.  It has been cited over 1200 times and  used to produce standard-weight curves for some species.  Recent work (Cole-Fletcher et al. 2011) suggests Carlander (1969) contains numerous errors in listed weight-length equations.  This paper assesses the weight-length relationships listed in Carlander for muskellunge, northern pike, and chain pickerel by comparing graphs of the weight vs. length equations with other data listed and with standard weight curves published by independent sources.  A number of discrepancies are identified through this analysis and new weight-length relationships are produced from listed data.

## Introduction

Weight-length relationships are a cornerstone of fishery research and management,  (Anderson and Neumann 1996) because of their importance in assessing body condition and estimating biomass from length class surveys. (Froese 2006)  Carlander's "Handbook of Freshwater Fishery Biology" (1969) includes key life history information on species in the pike genus (*Esox*) and many other important freshwater species.  Pike are of considerable economic and ecological importance. (Doyon et al. 1988)  Owing to some combination of compilation and unit conversion errors, the weight-length parameters tabulated in Carlander (1969) are not generally reliable and numerous errors have been propagated in the popular on-line database FishBase. (Cole-Fletcher et al. 2011)  This paper assesses the Carlander length-weight data in muskellunge (*E. masquinongy*), northern pike (E. *lucius*), and chain pickerel (E. *niger*) by graphing the weight-length relationships and comparing to the main data table also found in Carlander as well as the published standard weight curves (Anderson and Neumann 1996).  New weight-length relationships are generated in cases where Carlander (1969) includes applicable data, in the form $W(L) = 1000 \ (l/l_1)^b$ and in the traditional form $W(l) = al^b$, where $W$ is the weight in grams and $l$ is the total length in mm.   The relationship $W(l) = 1000 \ (l/l_1)^b$ is preferred, because $l_1$ is easily interpreted as the typical length of a fish weighing 1 kg, and because this relationship usually yields smaller uncertainties and covariances in the parameter values that result from regression analysis. (Dexter et al. 2011)

## Muskellunge (*Esox masquinongy*)

Figure 1a was created using the equations given on p. 355 of Carlander's "Handbook of Freshwater Fishery Biology" (1969).   Note the variation between the curves; H138 (East & Central Canada) suggests that 1000 mm long fish typically weigh over 16 kg, which is unrealistic for species with a long, narrow body type such as the muskellunge, and 198% of the standard weight.   The group of curves clustered near the standard weight curve seem reasonable, but the lowest curve, M329 ON central Canada muskies seems questionable at 74% of the standard weight curve.

Data in the table on pp. 356-358 of Carlander (1969) allows for a new weight-length relationship to be computed for muskellunge from the H138 data in east and central Canada.  This data is compared with both the new and errant weight-length relationship in Figure 1b.  The error bars in the figure were estimated as 20% of the mean weight divided by the square root of the number of samples.  Figure 1b makes it unmistakable that the weight-length relationship given on p. 355 of Carlander (1969) and repeated at FishBase.org (Froese and Pauly 2011) is in error. Rather than being significantly larger than the standard weight curve as suggested by the equation in Carlander (1969), the H138 weight-length data and best-fit curve is slightly below the standard weight curve, as expected for most weight-length relationships.

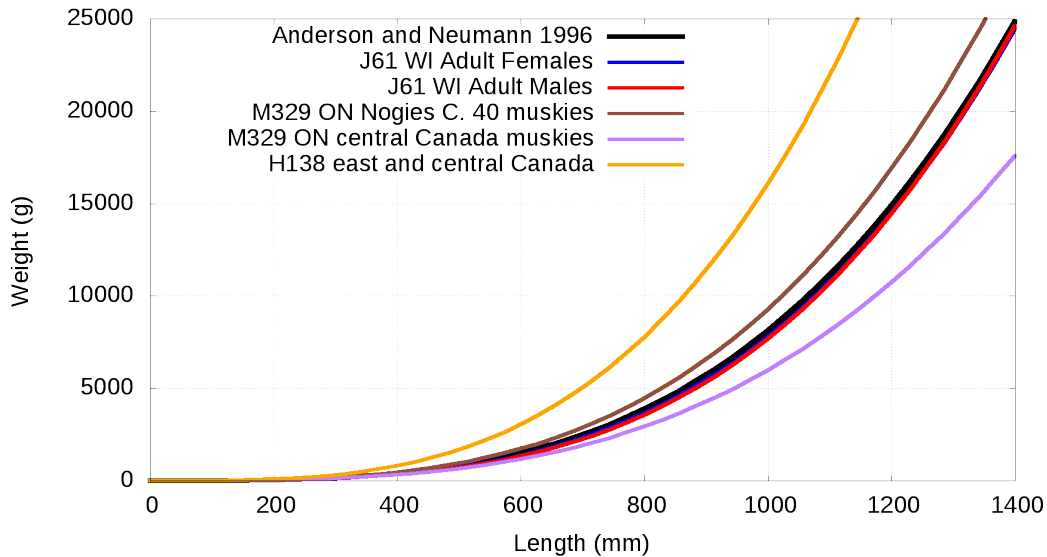

*Figure 1a: Curves from weight-length equations for muskellunge from Carlander (1969) p. 355 are compared with the standard weight curve. (Anderson and Neumann 1996) To facilitate comparing fork length relationships on the same graph, the the equation TL = 1.08 FL was used.*

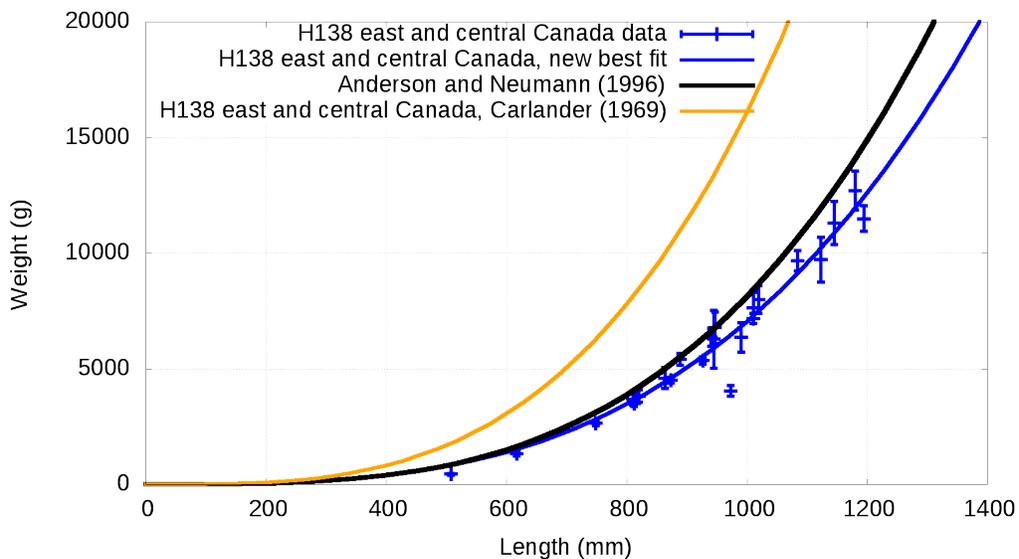

*Figure 1b: New best-fit curve for data from H138 east and central Canada from data in table from Carlander (1969) pp. 356-358 compared with errant weight-length equation from Carlander (1969) p. 355.*

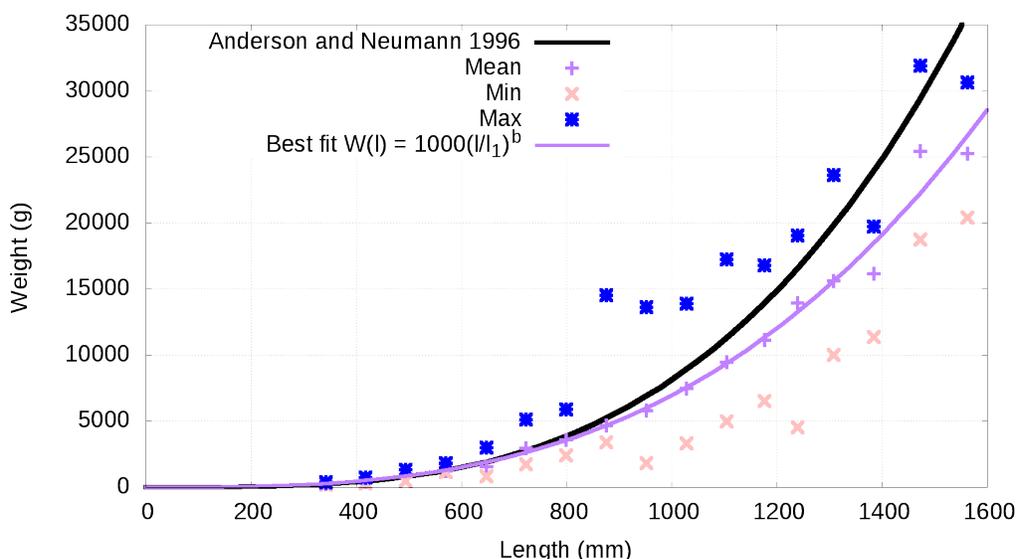

*Figure 2: Best fit curve of mean weights in main muskellunge weight-length data table in Carlander (1969) pp. 354-355. This plot assumes that the TL column in the Carlander table reports lengths in mm, in contrast to inches, which is the unit listed at the top of the table. Anderson and Neuman's standard weight curve is plotted also.*

The main muskellunge weight table given in Carlander (1969) pages 354-355 presents a new problem; the units on the total length (TL) column are obviously mislabeled. There is no chance a muskellunge could be 1600 inches, or 133.33 feet long. Carlander (1969) commonly lists total lengths in mm, especially in analogous tables for other species, so it is reasonable to assume that mm is intended here, and this interpretation yields consistent results, as shown in Figure 2.

It is reasonable that the best fit to the mean weights in the Carlander table is slightly below the standard weight curve, and that the standard weight curve is below most of the data for the maximum weight in most length classes, except for a few length classes with small sample sizes, since the standard weight represents an estimate of the 75[th] weight percentile, which might not be exceeded by the heaviest fish in a small sample. The parameter values from the best fit to the mean weights are shown in Table 1.

One of the populations of smaller muskellunge reported in Carlander (1969) was for George Lake, Wisconsin. Weight and length data for this population was reported from two sources (G152 and S499), and both sources suggest lower body condition and maximum length than most populations. Figure 3 shows the data and the best fit curve, and Table 1 reports the best fit parameters for this data. Figure 3 also shows the data reported by K126 on p. 360 of Carlander (1969), along with the best-fit curve.

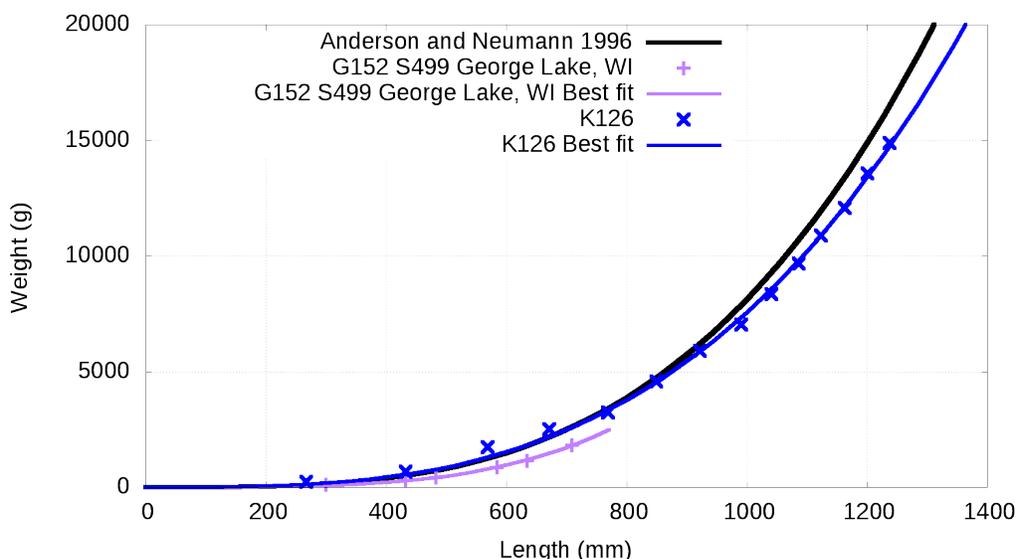

*Figure 3: Weight-length curves determined using data from from George Lake, WI data reported in Carlander (1969) pp. 356-358 from G152 and S499 and from K126 data reported on p. 360 along with Anderson and Neumann's standard weight curve.*

| | values | | | uncertainty | | |
|---|---|---|---|---|---|---|
| | **b** | **$l_1$** | **a** | **b** | **$l_1$** | **a** |
| **Mean** | 3.007 | 524.9 | 6.619E-06 | 5.2% | 5.1% | 112.2% |
| **H138** | 3.190 | 542.5 | 1.894E-06 | 5.7% | 3.9% | 126.2% |
| **G152/S499** | 3.798 | 608.1 | 2.665E-08 | 3.7% | 0.5% | 90.7% |
| **K126** | 3.132 | 524.3 | 3.038E-06 | 2.1% | 1.6% | 45.2% |

*Table 1: Best fit values for model $W(l) = 1000(l/l_1)^b$, for available Muskellunge data in Carlander (1969) for the Mean in the main weight-length table on pp. 354-355, the H138 data on pp. 356-359, the George Lake WI data from G152 and S499 on pp. 356-358, and the K126 data on p. 360.*

Table 1 shows reliable weight-length parameters obtained by non-linear least squares fitting to weight-length data available in Carlander (1969). The equivalent value of $a$ in the model $W(l) = al^b$, is computed as $a = 1000\ l_1^{-b}$. The uncertainties reported for $b$ and $l_1$ are those obtained by fitting to the improved model, $W(l) = 1000(l/l_1)^b$. The uncertainty reported for a is that obtained by directly fitting to the traditional model $W(l) = al^b$. The uncertainty in the values of $a$ reported here are actually much smaller, because these values of $a$ were inferred by fitting to the improved model; these uncertainties in $a$ are approximately $b$ times the relative uncertainty in $l_1$. For example the uncertainty in $a$ for K126 would be 2.1% times 3.132 = 6.6%, which is much smaller than the uncertainty in $a$ from fitting to the traditional model.

In summary, the weight-length relationships and data reported in Carlander (1969) contain several errors, including mislabeled units, and at least one weight-length curve that is well above the data from the same source. Our analysis provides several additional, reliable weight-length relationships along with uncertainties in the parameter estimates.

**Northern Pike (*Esox lucius*)**
To determine the accuracy of weight-length relationships in Carlander (1969), the weight-length equations on p. 337 are compared with each other and with the standard weight equation (Anderson and Neumann 1996) in Figure 4. Most of the curves are clustered around the standard weight curve; however, the C17 MN Lake of the Woods curve is so low (90% below at a total length of 1000 mm), that it is obviously in error. The B331 OH Lake Erie females curve is 19% below the standard weight curve at a total length of 1000 mm, which is not terribly unreasonable, but when combined with the fact that it is below the curve for B331 OH Lake Erie males for the entire range of adulthood, one might suppose that there is a probable error in this relationship as well.

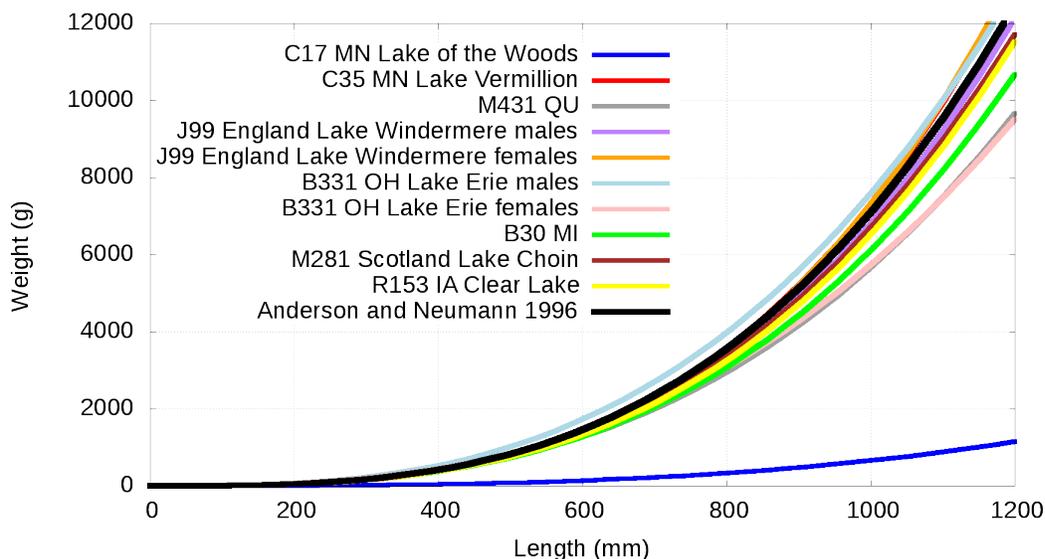

*Figure 4: Curves from weight-length equations for northern pike from Carlander (1969) p. 337 compared with the standard weight curve. (Anderson and Neumann 1996) To facilitate comparing fork length (FL), standard length (SL), and total length (TL) relationships on the same graph, the the equations TL = 1.06 FL and TL = 1.14 SL were used.*

Figure 5 shows the data for the mean, minimum, maximum, 25th, and 75th percentile weights for northern pike at lengths from the main weight-length table in Carlander (1969) pp. 336-337, along with the best fit curves for the mean, 25th, and 75th percentile weights compared with the standard weight curve. (Anderson and Neumann 1996) The closeness of the 75th percentile best fit with the standard weight curve gives confidence that both curves are reasonable; in fact, standard weight curves for some species were originally generated by fitting the 75th percentile data from a Carlander handbook. However, this approach often resulted in length-related biases in the standard weight curve, so standard practice moved to the regression line percentile approach instead. (Anderson and Neumann 1996) It is notable that the 75th percentile curve is significantly further above the mean best-fit curve than the 25th percentile curve is below it. In other words, the distribution of fish by weight is not even around the mean; it is skewed toward heavier weights. Being significantly below the mean weight is a significant impairment to survival and reproduction; whereas, being significantly above the mean weight is not. The best-fit parameters to the model $W(l) = 1000(l/l_1)^b$ are shown in Table 2.

Carlander (1969) contains data which are useful for determining a number of additional weight-length relationships which are shown in Figure 6. The data and best-fit equation for U11 SD Oahe L. is so close to the standard weight curve (Anderon and Neumann 1996), that they obscure each other. Most of the data and curves are below the standard weight curve, as expected, although the data for Athabaska lake (M26 as reported in Carlander 1969) is above the standard weight curve at the larger end of the length range.

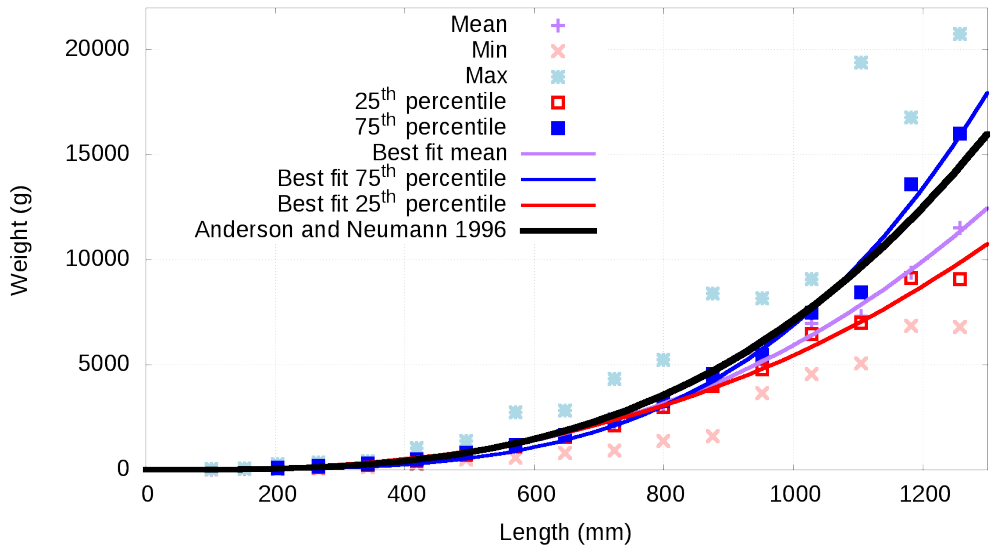

*Figure 5: Data and best-fit curves for main weight-length table for northern pike in Carlander (1969) pp. 336-337 compared with the standard weight curve. (Anderson and Neumann 1996)*

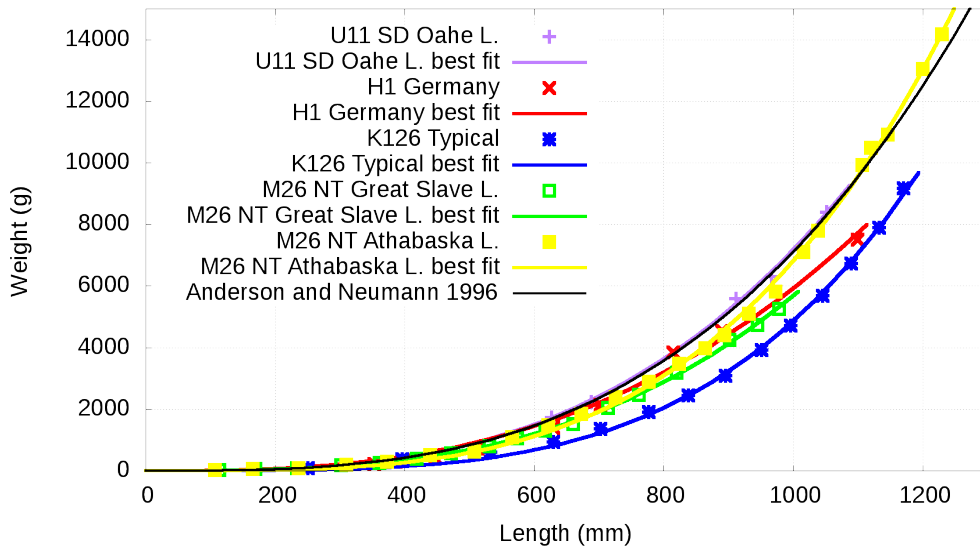

*Figure 6: Weight-length curves determined using data from from Carlander (1969): U11 and H1 data from pp. 341-343; K126 and M26 data from pp. 346-348. The standard weight curve (Anderson and Neumann, 1996) is plotted for comparison. Best fit parameters and uncertainties are given in Table 2.*

Table 2 shows reliable weight-length parameters obtained by non-linear least squares fitting to weight-length data available in Carlander (1969). The equivalent value of $a$ in the model $W(l) = al^b$, is computed as $a = 1000\ l_1^{-b}$. The uncertainties reported for $b$ and $l_1$ are those obtained by fitting to the improved model, $W(l) = 1000(l/l_1)^b$. The uncertainty reported for $a$ is that obtained by directly fitting to the traditional model $W(l) = al^b$. The uncertainty in the values of a reported here are actually much smaller, because these values of $a$ were inferred by fitting to the improved model. These uncertainties in $a$ are approximately $b$ times the relative uncertainty in $l_1$. For example the uncertainty in $a$ for U11 would be 3.5% times 3.061 = 10.7%, which is much smaller than the uncertainty in $a$ (73.3%) from fitting to the traditional model.

| | values | | | uncertainty | | |
|---|---|---|---|---|---|---|
| | *b* | *$l_1$* | *a* | *b* | *$l_1$* | *a* |
| **Mean** | 2.837 | 534.5 | 1.824E-05 | 2.7% | 2.0% | 53.0% |
| **75th Percentile** | 3.661 | 590.9 | 7.140E-08 | 4.8% | 3.3% | 122.7% |
| **25th Percentile** | 2.592 | 520.2 | 9.116E-05 | 5.0% | 3.8% | 90.0% |
| **U11 SD Oahe L.** | 3.061 | 525.6 | 4.696E-06 | 3.5% | 2.2% | 73.3% |
| **H1 Germany** | 2.771 | 526.4 | 2.873E-05 | 5.5% | 3.5% | 103.6% |
| **K126 Typical** | 3.893 | 666.4 | 1.017E-08 | 3.5% | 2.2% | 53.7% |
| **M62 NT Great Slave L.** | 3.048 | 565.8 | 4.071E-06 | 1.2% | 0.5% | 23.7% |
| **M62 NT Great Bear L.** | 3.155 | 550.8 | 2.254E-06 | 1.4% | 0.7% | 33.7% |
| **M62 NT Athabaska L.** | 3.551 | 582.7 | 1.511E-07 | 1.5% | 0.9% | 36.4% |

Table 2: Best fit values for model $W(l) = 1000(l/l_1)^b$, for available northern pike data in Carlander (1969) for the Mean , 75th, and 25th Percentile data in the main weight-length table on p. 337, the H1 and U11 data on pp. 341-344, and the K126 and M62 data on pp. 346-348.

In summary, the above analysis shows that one of the weight-length relationships reported in Carlander (1969) is in error and provides several additional, reliable weight-length relationships along with uncertainties in the parameter estimates. Performing a non-linear least-squares fit to the improved model, $W(l) = 1000(l/l_1)^b$, provides much smaller uncertainties in the parameter estimates than using the traditional model.

**Chain Pickerel (*Esox niger*)**
To determine the accuracy of weight-length relationships in Carlander (1969), the weight-length equations on p. 330 are compared with each other and with the standard weight equation (Anderson and Neumann 1996) in Figure 7. Both curves for the available equations are slightly below the standard weight curve, but not so low as to suspect an error. The equation from S186 was also compared with the original publication (Saila 1956) and found to be in agreement.

Figure 8 shows the data for the mean, minimum, maximum, 25th, and 75th percentile weights for chain pickerel from the main weight-length table in Carlander (1969) p. 330, along with the best fit curves for the mean, 25th, and 75th percentile weights compared with the standard weight curve. (Anderson and Neumann 1996) The closeness of the 75th percentile best fit with the standard weight curve gives confidence that both curves are reasonable. The 75th percentile curve is significantly further above the mean best-fit curve than the 25th percentile curve is below it, showing that the distribution of fish by weight is not even around the mean. For a given length; it is skewed toward the heavier weights. Being significantly below the mean weight is a significant impairment to survival and reproduction; whereas, being the same amount above the mean weight is not a significant impairment to survival and reproduction. The best-fit parameters to the model, $W(l) = 1000(l/l_1)^b$, are shown in Table 3.

Carlander (1969) contains data which are useful for determining a number of additional weight-length relationships shown in Figure 9. All three additional curves are slightly above the standard weight curve showing that these three populations were in excellent condition when the original measurements were made. This is not cause to doubt the standard weight curve since it was derived from 115 populations in 13 states. (Neumann and Flammang 1997) Table 3 shows reliable weight-length parameters obtained by non-linear least squares fitting to weight-length data available in Carlander (1969). The equivalent value of *a* in the model $W(l) = al^b$, is computed as $a = 1000 \, l_1^{-b}$. The uncertainties reported for *b* and $l_1$ are those obtained by fitting to the improved model, $W(l) = 1000(l/l_1)^b$. The uncertainty reported for *a* is that obtained by directly fitting to the traditional model $W(l) = al^b$. The uncertainty in the values of $a$ reported here are actually much smaller, because these values of $a$ were inferred by fitting to the improved model; these uncertainties in $a$ are approximately $b$ times the relative uncertainty in $l_1$.

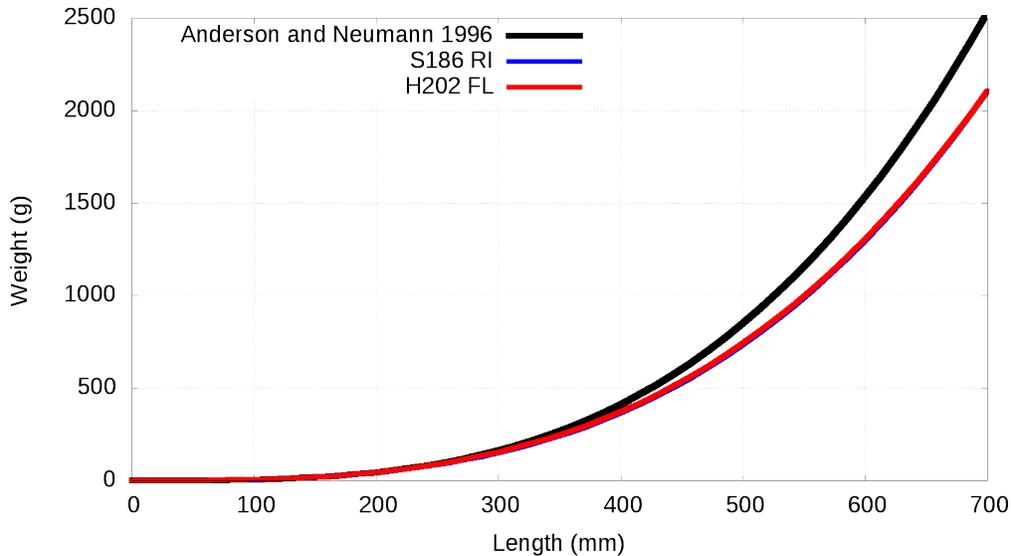

*Figure 7: Curves from weight-length equations for chain pickerel from Carlander (1969) p. 330 compared with the standard weight curve. (Anderson and Neumann 1996) To facilitate comparing fork length (FL) and total length (TL) relationships on the same graph, the the equations TL = 1.055 FL was used. The curve for H202 FL is very close to that for S186 RI, so that the curve for S186 RI is largely obscured.*

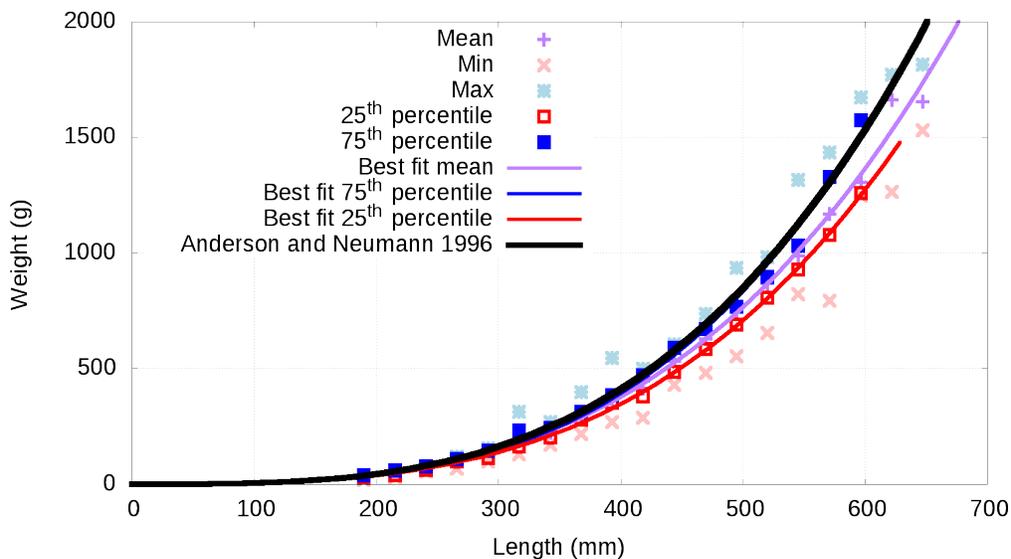

*Figure 8: Data and best-fit curves for main weight-length table for chain pickerel in Carlander (1969) p. 330 compared with the standard weight curve. (Anderson and Neumann 1996) The best-fit curve for the 75th percentile data is obscured by the standard weight curve.*

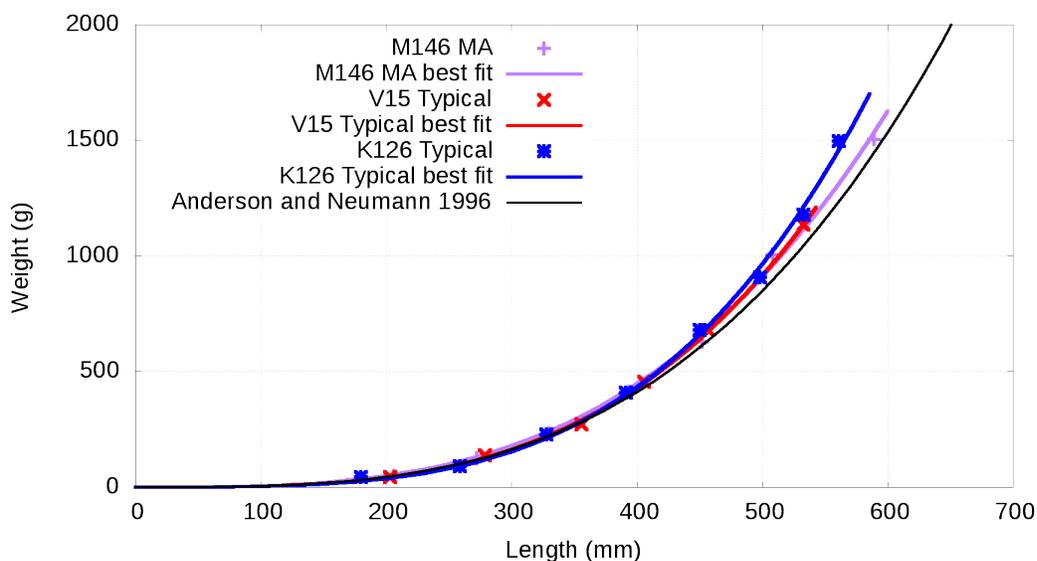

*Figure 9: Weight-length curves determined using data from from Carlander (1969):  M146  data from pp. 331-333;  V15 and K126 data from pp. 333-334.  The standard weight curve (Anderson and Neumann 1996) is also plotted for comparison.  Best fit parameters and uncertainties are given in Table 3.*

| | values | | | uncertainty | | |
|---|---|---|---|---|---|---|
| | $b$ | $l_1$ | $a$ | $b$ | $l_1$ | $a$ |
| **Mean** | 3.151 | 544.0 | 2.401E-06 | 2.6% | 0.4% | 52.3% |
| **75th Percentile** | 3.352 | 527.8 | 7.463E-07 | 3.1% | 0.4% | 63.8% |
| **25th Percentile** | 3.213 | 556.6 | 1.511E-06 | 1.1% | 0.2% | 21.5% |
| **M146 MA** | 3.190 | 515.3 | 2.225E-06 | 3.4% | 0.5% | 67.4% |
| **V15 Typical** | 3.395 | 513.4 | 6.271E-07 | 1.7% | 0.2% | 34.7% |
| **K126 Typical** | 3.596 | 505.3 | 1.903E-07 | 3.8% | 0.4% | 84.5% |

*Table 3: Best fit values for model $W(l) = 1000(l/l_1)^b$, for available chain pickerel data in Carlander (1969) for the Mean , 75th, and 25th Percentile data in the main weight-length table on p. 330, the M146 data on pp. 331-333, and the K126 and V15 data on pp. 333-334.*

In summary, the above analysis shows that the two weight-length equations reported for chain pickerel in Carlander (1969) are sound and provides several additional, reliable weight-length relationships along with uncertainties in the parameter estimates.  Performing a non-linear least-squares fit to the improved model,  $W(l) = 1000(l/l_1)^b$, provides much smaller uncertainties in the parameter estimates than using the traditional model.

**Conclusion**

This assessment revealed significant discrepancies in the weight-length relationships reported by Carlander (1969) in two of the three *Esox* species considered (muskellunge and northern pike).    Given that earlier work (Cole-Fletcher et al. 2011) has also cast doubt on weight-length relationships in Carlander (1969), it seems that Carlander's weight-length relationships need to be assessed carefully before being relied upon.    Since many of these weight-length relationships from Carlander are re-reported at Fishbase.org (Froese and Pauly 2011) whose error detection is not up to the task of noting all the glaring errors (Cole-Fletcher et al. 2011), the weight-length relationships reported at FishBase.org also need to be carefully assessed before being relied upon.

**Acknowledgements**
This work was funded, in part, by BTG Research (www.btgresearch.org).  The authors are grateful to several reviewers for carefully reading the manuscript and offering constructive suggestions.